\documentclass[prd,aps,preprintnumbers,nofootinbib,twocolumn]{revtex4}
\usepackage{epsfig}
\usepackage{amsmath}
\usepackage{hyperref}
\usepackage{multirow}
\usepackage[utf8]{inputenc}

\def\as{\alpha_S}
\def\nf{n_f}

\begin{document}

\preprint{Cavendish--HEP--19/12}

\renewcommand{\thefigure}{\arabic{figure}}

\title{Ambiguities of the principle of maximum conformality procedure for hadron collider processes}

\author{Herschel A. Chawdhry}
\author{Alexander Mitov}
\affiliation{{\small Cavendish Laboratory, University of Cambridge, Cambridge CB3 0HE, United Kingdom}}

\date{\today}

\begin{abstract}
In any calculation in perturbative Quantum Chromodynamics (QCD) a choice needs to be made for the unphysical renormalisation scale, $\mu_R$. The Brodsky-Lepage-Mackenzie/Principle of Maximum Conformality (BLM/PMC) scale-setting procedure is one proposed method for selecting this scale. In this work we identify three ambiguities in the BLM/PMC procedure itself. Their numerical impact is studied using the example of the total cross-section for $t\bar{t}$ production through Next-to-Next-to-Leading Order in QCD. One ambiguity is the arbitrary choice of the value of the highest-order PMC scale. The numerical impact of this choice on the BLM/PMC prediction for the cross-section is found to be comparable to the impact of the choice of $\mu_R$ in the conventional scale-setting approach. Another ambiguity relates to the definitions of the other PMC scales and is similarly found to have a large impact on the cross-section.
\end{abstract}
\maketitle

\section{Introduction\label{sec:intro}}

When performing calculations in Quantum Chromodynamics (QCD), any partonic observable $\rho$ is usually calculated as a perturbative series in the strong coupling constant $\as$: 
\begin{equation}\label{eq:rho}
\rho = \sum\limits_{n}c_n(\mu_R)\left(\frac{\as(\mu_R)}{4\pi}\right)^n.
\end{equation}
The renormalization scale $\mu_R$ is an arbitrary parameter which enters this equation following renormalization. Formally, when working to all orders in $\as$, the $\mu_R$-dependence of the coefficients $c_n$ exactly compensates that of $\as$ so that $\rho$ is independent of $\mu_R$. In practice, however, the perturbative series is truncated beyond some finite order, $N$, and this causes $\rho$ to become $\mu_R$-dependent.

Conventionally, in processes with a single hard scale $Q$, one chooses $\mu_R=Q$ on dimensional grounds. Although not essential for the goals of the present work, we would like to mention that more refined arguments for choosing this scale have been given in the literature \cite{Nason:1989zy,Beenakker:1990maa,Mangano:1991jk,Catani:2001cc,Maltoni:2003pn,Boos:2003yi,Alwall:2007fs,Bauer:2009km,Melnikov:2009wh,Ahrens:2010zv,Berger:2010zx,Denner:2012yc,Anastasiou:2016cez,Boughezal:2016yfp,Czakon:2016dgf,Currie:2017eqf,Czakon:2018nun}. Such arguments tend to modify the choice $\mu_R=Q$ by a factor of ${\cal O}(1)$ and are especially relevant for observables with several kinematic scales. 

The value of $\mu_R$ is then varied in a range $(Q/2, 2Q)$ and the resulting variation in the value of $\rho$ is taken to be representative of the error which arises from omitting the $\mathcal{O}\left(\as^{N+1}\right)$ terms from Eq.~(\ref{eq:rho}). While the choice of this variation range is a matter of convention, its adequacy is justified {\it a posteriori} by higher-order calculations.

The Brodsky-Lepage-Mackenzie/Principle of Maximum Conformality (BLM/PMC) method \cite{Brodsky:1982gc,Brodsky:2011ta,Brodsky:2011ig} has been proposed as a way of removing the renormalization scale ambiguity. The method is based on an appealing physical motivation and, as explained in Section \ref{sec:BLM/PMC}, it algorithmically prescribes a ``correct'' value for the scale $\mu_R$. The method has been applied to a number of processes including Higgs production \cite{Spira:1995rr}, meson production \cite{Belitsky:2001nq, Ivanov:2004zv, Anikin:2004jb, Belitsky:2005qn}, pion form factors \cite{Brodsky:1997dh, Bakulev:2000uh}, $b$-physics \cite{Voloshin:1995sf, Bigi:1997fj}, and $t\bar{t}$ production \cite{Brodsky:2012rj,Brodsky:2012sz,Brodsky:2012ik,Wang:2014sua,Wang:2015lna,Wang:2017kyd}. Some possible generalisations of the BLM/PMC method have been discussed in Refs. \cite{Kataev:2014jba,Kataev:2014zwa,Kataev:2016aib,Aleshin:2019yqj,Garkusha:2018mua}.

In this paper we address the following question: {\it are there any ambiguities associated with the BLM/PMC method and what is their numerical impact?}

Since scale variations are usually interpreted as representing theory uncertainties, the BLM/PMC method might appear to eliminate uncertainties in theoretical predictions. In Sections \ref{sec:BLM/PMC} and \ref{sec:resultsdiscussion}, we will discuss the extent to which this is true.

In order to keep our discussion less abstract we will consider the process of top-pair production at hadron colliders, which is well-suited for this study given that it is fully known through NNLO, has generic kinematics and color structure, and is very precisely measured. The application of the BLM/PMC method to this process has been extensively studied \cite{Brodsky:2012rj,Brodsky:2012sz,Brodsky:2012ik,Wang:2014sua,Wang:2015lna,Wang:2017kyd}. We expect that many of our findings transcend this particular process.

\section{Background: The BLM/PMC procedure at NNLO}\label{sec:BLM/PMC}

One applies the BLM/PMC method to a partonic observable $\rho$ like the one in Eq.~(\ref{eq:rho}). If this is a hadron collider observable, as we would consider in this paper, then two qualifications are required.

Firstly, in order to construct the proper hadron-level observable, Eq.~(\ref{eq:rho}) needs to be convolved with parton distribution functions (PDFs) and summed over all possible initial partonic states. Such partonic observables are not uniquely defined since they depend on the scheme used to subtract collinear singularities; we will not be concerned with this here and will assume a given factorization scheme (the ${\overline {\rm MS}}$ scheme is standard). 

Secondly, the perturbative coefficients $c_n$ also depend on the unphysical factorization scale $\mu_F$, which separates the long-distance physics absorbed into the PDFs from the short-distance physics in the  perturbative coefficients $c_n$. The BLM/PMC method does not prescribe a value for $\mu_F$. In this work we will focus exclusively on the scale $\mu_R$ and will fix the factorization scale at some standard value, as was also done in the previous BLM/PMC work on the subject \cite{Brodsky:2012rj,Brodsky:2012sz,Brodsky:2012ik,Wang:2014sua,Wang:2015lna,Wang:2017kyd}. For the total top-pair cross-section this is $\mu_F=m_t$ (although a smaller value  $\mu_F=m_t/2$ may be more appropriate \cite{Czakon:2016dgf}). In the following we will suppress the explicit dependence of the coefficients $c_n$ and the observable $\rho$ on the partonic channel and factorization scale. 

The idea behind the BLM/PMC method is to first identify the terms proportional to the QCD $\beta$-function coefficients $\beta_i$ inside the partonic coefficients $c_n$ and absorb them into the running coupling by making a suitable choice for the renormalization scale $\mu_R$. 

At Next-to-Leading Order (NLO) in QCD, one can use the BLM method \cite{Brodsky:1982gc} and, for any given process, uniquely fix the value of $\mu_R$ by requiring the LO and NLO perturbative coefficients $c_n$ to be independent of $\beta_i$. As it turns out, however, beyond NLO one cannot absorb all $\beta_i$ coefficients into the running coupling with a single choice of scale. The PMC method \cite{Brodsky:2013vpa} extends the BLM idea to higher-order QCD calculations by using a different value for the renormalization scale at each order in $\as$. After such a choice the partonic observable $\rho$ in Eq.~(\ref{eq:rho}) takes the form:
\begin{equation}\label{eq:rho_PMC}
\rho = \sum\limits_{n}\tilde c_n\left(\frac{\as(q_n)}{4\pi}\right)^n,
\end{equation}
for some new coefficients $\tilde c_n$ and scales $q_n$. These new scales and coefficients are chosen by requiring that
\begin{enumerate}
\item the coefficients $\tilde c_n$ are independent of $\beta_i$\,, and
\item Eqs.~(\ref{eq:rho}) and (\ref{eq:rho_PMC}) agree through order $\as^N$\,.
\end{enumerate}

It is convenient to express the coefficients $c_n$ in Eq.~(\ref{eq:rho}) through a new set of $\beta_i$-independent coefficients $s_{n,k}(\mu_R)$. As mentioned above, in the rest of this paper we will specialize our discussion to the inclusive cross-section for top-quark pair production. This means that the sum in Eqs.~(\ref{eq:rho}) and (\ref{eq:rho_PMC}) goes from $n=2$ (the LO term) through $n=N=4$ (the NNLO term). In this context the coefficients $s_{n,k}(\mu_R)$ are defined by means of the following implicit equations:
\begin{eqnarray}
c_2 &=& s_{2,0} \,, \nonumber\\
c_3 &=& s_{3,0} + 2s_{3,1}\beta_0\,, \nonumber\\
c_4 &=& s_{4,0} + 2s_{3,1}\beta_1 + 3s_{4,1}\beta_0 + 3s_{4,2}\beta_0^2\,.
\end{eqnarray}

The $\mu_R$-dependence of the coefficients $s_{n,k}$ follows from the requirement that observables are independent of $\mu_R$. In particular, one finds that $s_{n,0}$ have no dependence on $\mu_R$.

We remark on a practical aspect of the procedure outlined above. The $\beta_i$ dependence is inferred from the known $\nf$ dependence of the cross-section by inverting the dependence of $\beta_i$ on $\nf$:
\begin{equation}
\beta_0 = 11 - \frac{2}{3}\nf \,,~~\beta_1 = 102 - \frac{38}{3}\nf \,.
\label{eq:beta01}
\end{equation}
The above procedure requires the exclusion of $\nf$ contributions from light-by-light type of diagrams that are not associated with coupling renormalization. In the process at hand, no light-by-light contribution is present in the $q\bar q$ initiated contribution. The $gg$ initiated one does contain such diagrams at NNLO but these contributions have not been separated in the existing literature. We thus neglect to separate them in this work. To the best of our knowledge they have likewise not been separately accounted for in the previous applications of the BLM/PMC method to top-quark pair production. 

The PMC coefficients $\tilde c_n$ and scales $q_n$ are defined by
\begin{eqnarray}
\tilde c_n &=& s_{n,0} \,, \label{eq:cn}\\
\log\left(\frac{q_2^2}{\mu_0^2}\right) &=& -\frac{s_{3,1}}{s_{2,0}} + \frac{3}{2}\left[\left(\frac{s_{3,1}}{s_{2,0}}\right)^2-\frac{s_{4,2}}{s_{2,0}}\right]\beta_0\frac{\as}{4\pi}\,, \label{eq:logQ2}\\
\log\left(\frac{q_3^2}{\mu_0^2}\right) &=& -\frac{s_{4,1}}{s_{3,0}} \,, \label{eq:logQ3}
\end{eqnarray}
where $s_{n,k}=s_{n,k}(\mu_0)$ and $\as=\as(\mu_*)$ in Eq.~(\ref{eq:logQ2}).

Two new scales, $\mu_0$ and $\mu_*$, appear in Eqs.~(\ref{eq:logQ2},\ref{eq:logQ3}). The $\mu_0$-dependence of Eqs.~(\ref{eq:logQ2},\ref{eq:logQ3}) is purely formal: it can be shown that the scales $q_2$ and $q_3$ are completely independent of $\mu_0$. In other words, the $\mu_0$ dependence of the functions $s_{n,k}$ is such that all $\mu_0$ dependence in Eqs.~(\ref{eq:logQ2},\ref{eq:logQ3}) cancels between the two sides of those equations.

Eq.~(\ref{eq:logQ2}) also depends on the scale $\mu_*$, whose value is arbitrary. This is so since a change in $\mu_*$ only affects the relation $c_n\leftrightarrow \tilde c_n$ with terms beyond NNLO. 

The term in the square bracket in Eq.~(\ref{eq:logQ2}) vanishes for observables that respect the so-called large-$\beta_0$ approximation. As follows from Refs.~\cite{Baernreuther:2012ws,Czakon:2012zr,Czakon:2012pz,Czakon:2013goa} this is not the case for top-quark pair production. In the $q\bar{q}\,$ partonic reaction (introduced in Sec.~\ref{sec:method} below) the square bracket term is a pure number (see the related discussion in Ref.~\cite{Czakon:2014xsa}) while the corresponding result for the $gg\,$ reaction is only known as a precise numeric fit \cite{Czakon:2013goa}.

Following Ref.~\citep{Brodsky:2012sz}, one also needs to subtract the so-called ``Coulomb" terms from all functions $s_{n,k}$ that enter Eqs.~(\ref{eq:logQ2},\ref{eq:logQ3}) for all partonic reactions that contain such terms. The subtraction procedure of the Coulomb terms in top-quark pair production is explained in detail in Sec.~\ref{sec:method} below.

Clearly the choice of the scale $\mu_*$ does have an impact on the values of the scales $q_n$ and this represents one ambiguity in the PMC procedure. We find its numerical impact to be small and we suspect it is responsible for the very small ``initial renormalization scale dependence'' reported in Ref.~\cite{Brodsky:2012ik}. When presenting numerical results in Section \ref{sec:resultsdiscussion}, we will therefore focus on two other ambiguities, which we will now describe and whose numerical impact is larger.

We note that the PMC scales $q_n$ defined above are not the only way to absorb the $\beta_i$-dependence of the coefficients $c_n$ into the running coupling. For example, one could modify Eqs.~(\ref{eq:logQ2},\ref{eq:logQ3}) by defining an alternative set of scales $q'_n$:
\begin{eqnarray}
\log\left(\frac{q_2^{\prime 2}}{\mu_R^2}\right) &=& -\frac{s_{3,1}}{s_{2,0}} \,, \label{eq:logQ2prime}\\
\log\left(\frac{q_3^{\prime 2}}{\mu_R^2}\right) &=& -\frac{s_{4,1}}{s_{3,0}} + \frac{s_{2,0}}{s_{3,0}}\left[\left(\frac{s_{3,1}}{s_{2,0}}\right)^2-\frac{s_{4,2}}{s_{2,0}}\right]\beta_0 \,, \label{eq:logQ3prime}
\end{eqnarray}
which have the advantage of not containing the arbitrary scale $\mu_*$.
Clearly, the choice of whether to work with the scales $q_n$ or $q'_n$ represents a second ambiguity in the application of the PMC procedure.\footnote{Arguments in favour of using $q_n$ have been given in Ref. \cite{Brodsky:2013vpa}.} In what follows, we carry out our calculations using the original scales $q_n$ as well as the alternative scales $q'_n$ and explore the numerical difference between the two.

The last PMC scale, $q_4$, which appears at NNLO remains arbitrary at this order. Its fixing requires the knowledge of the $\beta_i$-dependent terms in the ${\rm N}^3{\rm LO}$ coefficient functions for $t\bar t$ production. These are not available at present. The arbitrariness of the scale $q_4$ represents the third, and most significant, ambiguity which we have identified in the PMC procedure. In Ref.~\cite{Brodsky:2013vpa}, the choice is made to set it equal to the previous known scale, $q_3$. While this is a plausible choice, we are not aware of a motivation in its favour. In what follows, in order to illustrate the significance of this ambiguity, we explore two choices: $q_4=q_3$ and $q_4=m_t$, where $m_t$ is the (pole) mass of the top quark. 

We wish to make one remark on the subject of theoretical uncertainties. The BLM/PMC framework asserts that there is a unique ``correct'' way of choosing the renormalisation scale, and that one should not try to estimate theoretical uncertainties by varying this scale in the manner described in Section \ref{sec:intro}. Nevertheless, we emphasise that the ``renormalisation-scale uncertainty'' conventionally quoted in perturbative QCD predictions is only a proxy for the error arising from the truncation of the sum in Eq.~(\ref{eq:rho}). Prescribing a procedure to choose $\mu_R$ may remove the way to estimate this error but it cannot remove the error itself, even in the absence of any ambiguities in the scale-setting procedure.

\section{Details about the implementation}\label{sec:method}

All partonic contributions to the total inclusive NNLO cross-section for $t\bar{t}$ production have been calculated in Refs.~\cite{Baernreuther:2012ws,Czakon:2012zr,Czakon:2012pz,Czakon:2013goa}, keeping their $\nf$-dependence explicit. As explained above Eq.~(\ref{eq:beta01}), we convert this $\nf$-dependence into a dependence on the coefficients $\beta_i$. The factorization scale is set to $m_t$ in all partonic reactions. The value of the renormalization scale for each partonic reaction is different, according to what is prescribed for it by the PMC approach. In fact we apply the PMC procedure only to the two dominant partonic channels $gg\rightarrow t\bar{t}+X$ and $q\bar{q}\rightarrow t\bar{t}+X$. All other contributing partonic reactions are included, as appropriate, only in the predictions for the complete hadron-level cross-section. For these sub-dominant channels, the standard choice $\mu_R=\mu_F=m_t$ is made.

In addition to depending on $\mu_R$ and $\mu_F$, the partonic cross-section coefficients $c_n$ also depend on $m_t$ and the partonic center of mass energy $\hat s$ through the following variable:
\begin{equation}
v=\sqrt{1-4m_t^2/\hat{s}}\,.
\label{eq:v}
\end{equation}

As mentioned in the previous section, in order to derive the PMC scales $q_2, q_3$ (or $q'_2, q'_3$), in each of the $q\bar q$ and $gg$ partonic reactions we first subtract the ``Coulomb" terms from the functions $s_{3,0}$ and $s_{4,1}$ (their explicit expressions can be found in Ref.~\cite{Beneke:2009ye}). The Coulomb terms in the function $s_{4,0}$ are not subtracted since they do not enter the scales $q_2, q_3$ (or $q'_2, q'_3$) through NNLO. The Coulomb terms are identified as the terms proportional to $1/v$ or $\log(v)/v$ in the series expansion of the functions $s_{3,0}/v$ and $s_{4,1}/v$ around $v=0$. The subtracted Coulomb contributions include terms $\sim\log\left({\mu_R}\right)$, as appropriate.

We only apply PMC scale-setting to the remaining part of the partonic cross-section. The Coulomb terms are then added back. Since they constitute only a small part of the partonic cross-section, we do not apply the PMC procedure to the Coulomb terms themselves. 

The reason for the separate treatment of the Coulomb terms is that at sufficiently high orders, the integrability of the cross-section requires their factorization into a toponium-like wave-function. A detailed analysis can be found in Ref.~\cite{Beneke:2016jpx}. 

We find that the subtraction of the Coulomb terms has a large impact on the PMC scales $q_2$ and $q_3$: in fact, failure to subtract the Coulomb terms leads to $q_3 \ll \Lambda_{\textrm{QCD}}$ and hence a divergent cross-section.

Finally, in our numerical predictions for the hadronic $t\bar t$ cross-section we use the PDF set {\tt NNPDF3.1} \cite{Ball:2017nwa} and we set $m_t=173.3$ GeV. We have verified that the PDF set {\tt CT14} \cite{Dulat:2015mca} produces similar results to those shown here. We base our numerical calculations on a modified version of the program {\tt Top++} \cite{Czakon:2011xx}.

\section{Results and Discussion\label{sec:resultsdiscussion}}

\subsection{PMC scales for $gg$ and $q\bar{q}$ channels}

\begin{figure}[h] 
\includegraphics[width=\linewidth,trim=0 3mm 0 1mm]{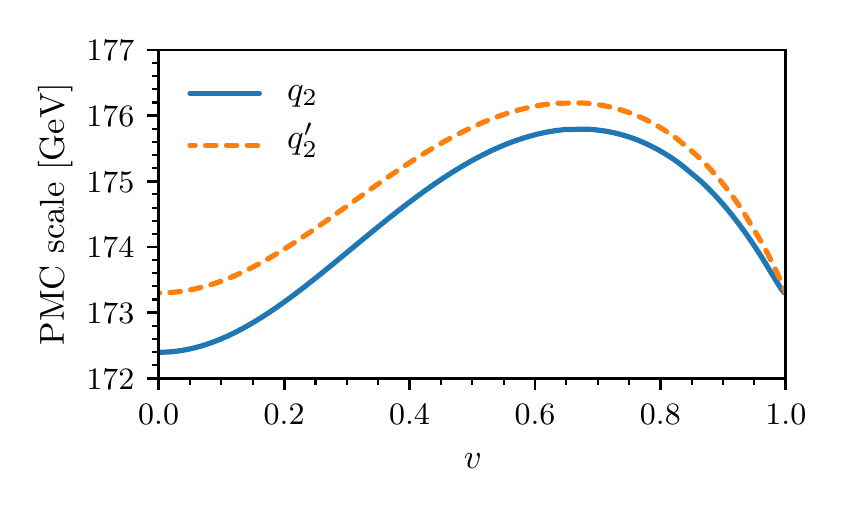}
\includegraphics[width=\linewidth,trim=0 5mm 0 10mm]{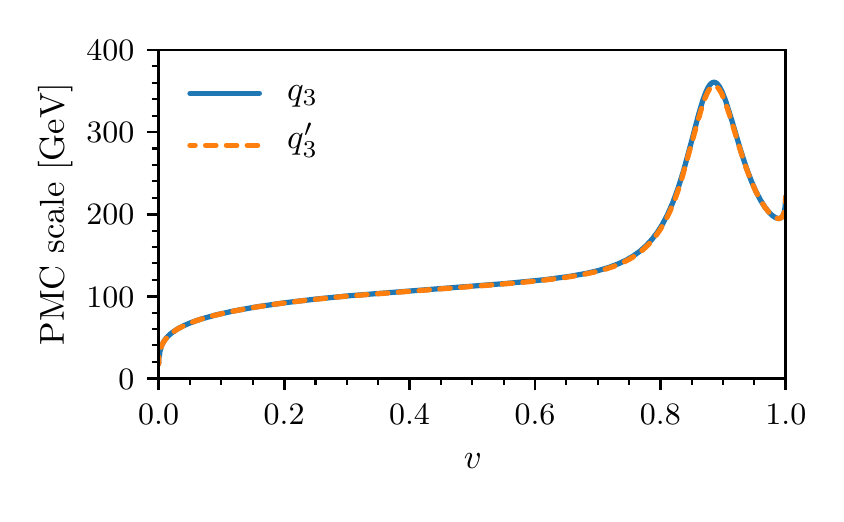}
\caption{The PMC scales $q_2$, $q'_2$, $q_3$, $q'_3$ for the $gg$ channel as functions of the relative velocity of the final-state top quarks.}
\label{fig:gg_q23}
\end{figure}

\begin{figure}[h]
\includegraphics[width=\linewidth,trim=0 3mm 0 1mm]{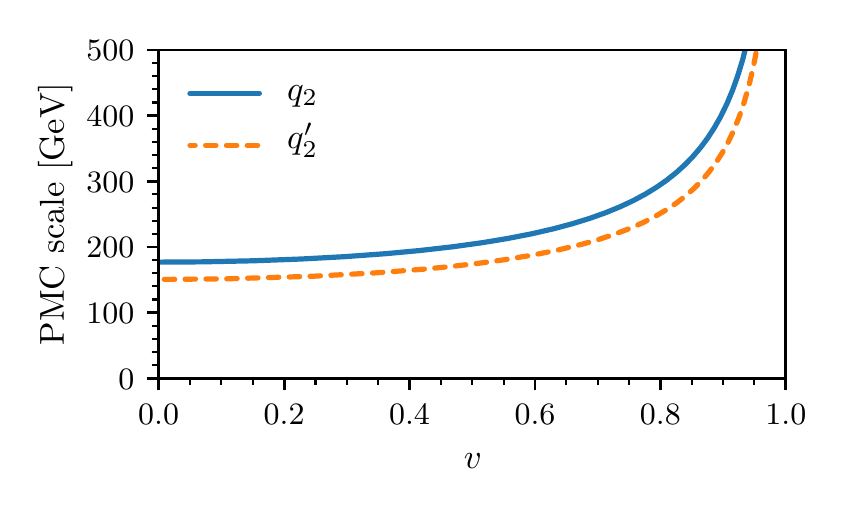}
\includegraphics[width=\linewidth,trim=0 5mm 0 10mm]{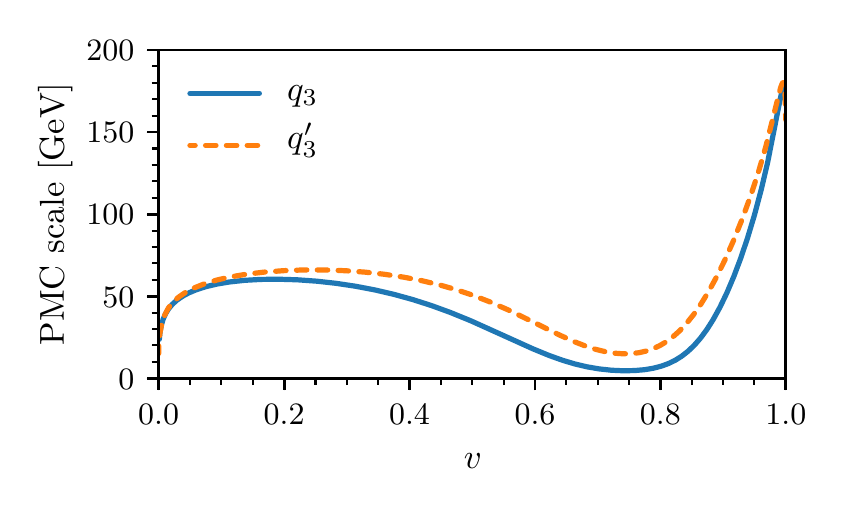}
\caption{As in Fig.~\ref{fig:gg_q23} but for the $q\bar{q}$ channel.}
\label{fig:qqbar_q23}
\end{figure}

Applying the formulae from Sec.~\ref{sec:BLM/PMC}, we now derive the PMC scales for the $gg$ and $q\bar q$ channels. We remind the reader that these scales depend on the parton level kinematics (through the variable $v$ for the case of the inclusive $t\bar{t}$ cross-section) but are independent of PDFs and, by extension, of the type of collider ($pp$ versus $p\bar p$) or collider energy. The results are shown in Fig.~\ref{fig:gg_q23} and \ref{fig:qqbar_q23} for, correspondingly, the $gg$ and $q\bar q$ channels.


For the $q\bar{q}$ channel (Fig.~\ref{fig:qqbar_q23}), it is interesting to observe that in the kinematic region $v \in [0.7,0.8]$, the scale $q_3$ reaches values as low as 4.6 GeV.
In fact, without the Coulomb subtraction procedure outlined in Sec.~\ref{sec:method}, $q_3$ takes values below $10^{-10}$ GeV in this kinematic region. Similar singularities have previously been found in vector meson production \cite{Anikin:2004jb}.

\subsection{Cross-sections for $t\bar{t}$ production at LHC13 and Tevatron} \label{subsec:results_crosssections}

Having derived the PMC scales for the $gg$ and $q\bar{q}$ channels, we will now calculate hadron-level cross sections at the 13 TeV LHC and also at the Tevatron. For each collider, we will compare the results from PMC scale-setting to those from the conventional approach ($\mu_R = m_t$). For the latter, uncertainties are computed by varying $\mu_R$ in the range $(m_t/2, 2m_t)$. 
\footnote{No numerical estimate is made for the theoretical uncertainty in the BLM/PMC predictions, as explained at the end of Sec.~\ref{sec:BLM/PMC}.} 
We remind the reader that throughout this work we have fixed $\mu_F=m_t$.

In this section, we use the ``standard'' choice of PMC scales, i.e. the scales $q_2$ and $q_3$ as defined in Eqns.~(\ref{eq:logQ2}) and~(\ref{eq:logQ3}), and setting $q_4=q_3$. The effect of alternative choices will be explored in Sec.~\ref{subsec:resultsdiscussion_ambiguities}.

At the 13 TeV LHC, using the BLM/PMC method, we obtain the following prediction for the total hadron-level cross-section for $pp\rightarrow t\bar{t}+X$:
\begin{equation}
\sigma_{\textrm{BLM/PMC}} = 813 \textrm{ pb}\,.
\end{equation}
For comparison, the predicted cross-section using conventional scale-setting is:
\begin{equation}
\sigma_{\textrm{Conventional}} = 794^{+28}_{-39} \textrm{ pb}\,,
\end{equation}
and the most recent precise experimentally-measured values from ATLAS \cite{Aaboud:2016pbd} and CMS \cite{Sirunyan:2018goh} are:
\begin{eqnarray}
\sigma_{\textrm{ATLAS}} &=& 818 \pm 8 \pm 27 \pm 19 \pm 12 \textrm{ pb}\,,\\
\sigma_{\textrm{CMS}} &=& 803 \pm 2 \pm 25 \pm 20 \textrm{ pb}\,.
\end{eqnarray}

\begin{table}[]
\begin{tabular}{|c|r|r|r|r|}
\hline
\multirow{2}{*}{}     & \multicolumn{2}{c|}{$q\bar{q}$ channel}               & \multicolumn{2}{c|}{$gg$ channel}                    \\ \cline{2-5}
                      & \multicolumn{1}{c|}{~\,PMC~~} & \multicolumn{1}{c|}{Conv.} & \multicolumn{1}{c|}{~\,PMC~~} & \multicolumn{1}{c|}{Conv.} \\ \hline
$\alpha_s^2$ {[}pb{]} & $62.4$ & $68.5$ & $405.7$ & $406.9$ \\
$\alpha_s^3$ {[}pb{]} & $41.7$ & $8.5$ & $256.4$ & $220.8$ \\
$\alpha_s^4$ {[}pb{]} & $-32.3$ & $4.7$ & $76.4$ & $81.5$ \\ \hline
NNLO {[}pb{]} & $71.8$ & $81.8^{+1.9}_{-2.2}$ & $738.4$ & $709.2^{+28.1}_{-37.2}$ \\ \hline
\end{tabular}
\caption{Contribution of the $q\bar{q}$ and $gg$ channels to $\sigma_{pp\to t\bar{t}+X}$ at the 13 TeV LHC at each order in $\alpha_s$.}
\label{tab:sigmaLHC13}
\end{table}
\begin{table}[]
\begin{tabular}{|c|r|r|r|r|}
\hline
\multirow{2}{*}{}     & \multicolumn{2}{c|}{$q\bar{q}$ channel}               & \multicolumn{2}{c|}{$gg$ channel}                    \\ \cline{2-5}
                      & \multicolumn{1}{c|}{~\,PMC~~} & \multicolumn{1}{c|}{Conv.} & \multicolumn{1}{c|}{~\,PMC~~} & \multicolumn{1}{c|}{Conv.} \\ \hline
$\alpha_s^2$ {[}pb{]} & $4.55$ & $4.89$ & $0.39$ & $0.39$ \\
$\alpha_s^3$ {[}pb{]} & $3.31$ & $0.96$ & $0.41$ & $0.33$ \\
$\alpha_s^4$ {[}pb{]} & $-2.24$ & $0.42$ & $0.19$ & $0.18$ \\ \hline
NNLO {[}pb{]} & $5.62$ & $6.27^{+0.16}_{-0.20}$ & $0.98$ & $0.91^{+0.07}_{-0.07}$ \\ \hline
\end{tabular}
\caption{As in Table~\ref{tab:sigmaLHC13} but for $\sigma_{p\bar{p}\to t\bar{t}+X}$ at the Tevatron.}
\label{tab:sigmaTevatron}
\end{table}

At the Tevatron, we find the following BLM/PMC prediction for the total $p\bar{p} \rightarrow t\bar{t}+X$ cross-section:
\begin{equation}
\sigma_{\textrm{BLM/PMC}} = 6.48 \textrm{ pb}\,.
\end{equation}
For comparison, the cross-section using conventional scale-setting is:
\begin{equation}
\sigma_{\textrm{Conventional}} = 7.06^{+0.21}_{-0.25} \textrm{ pb}\,,
\end{equation}
and the experimentally-measured value \cite{Aaltonen:2013wca}
is:
\begin{equation}
\sigma_{\textrm{Experimental}} = 7.60 \pm 0.41 \textrm{ pb}\,.
\end{equation}

To examine the origin of these values, the contributions of the two dominant partonic channels ($gg$ and $q\bar{q}$) to these cross-sections are shown in Tables~\ref{tab:sigmaLHC13} (for the LHC) and~\ref{tab:sigmaTevatron} (for the Tevatron). In each case, a breakdown is provided, showing the contributions from each power of $\as$.

In both tables it can be seen that the BLM/PMC procedure leads to a slower convergence than in conventional scale-setting. Similar behaviour has previously been discussed in Refs. \cite{Grunberg:1991ac,Kataev:2014jba,Cvetic:2016rot}.

\subsection{Effect of ambiguities}\label{subsec:resultsdiscussion_ambiguities}

We next explore the effects of the ambiguities in the BLM/PMC procedure which were outlined in Sec.~\ref{sec:BLM/PMC}. In order to do so, we recompute the above cross-sections using a variety of choices for the PMC scales ($q_2, q_3, q_4$):
\begin{enumerate}
\item ($q_2, q_3, q_3$) Firstly, we restate the results using the ``standard'' choice of PMC scales which were used in the previous section.
\item ($q_2, q_3, m_t$) Secondly, we study the numerical impact of the arbitrary choice of the scale $q_4$ by setting $q_4=m_t$ rather than $q_4=q_3$.
\item ($q'_2, q'_3, q'_3$) Thirdly, we explore the other main ambiguity discussed in Sec.~\ref{sec:BLM/PMC}, by using the scales $q'_n$ rather than $q_n$.
\item ($m_t$) For the purposes of comparison, we also present the results using the conventional choice $\mu_R=m_t$.
\end{enumerate}

The contribution of the $gg$ channel to the 13 TeV LHC cross-section, as predicted by each of these choices of scales, is shown in Table~\ref{tab:ambiguities_gg}. Similarly, the contribution of the $q\bar{q}$ channel is shown in Table~\ref{tab:ambiguities_qqbar}. The total cross-section, incorporating the contributions from all partonic channels, is shown in Table~\ref{table:NNLO_full_cross-sections}, where alongside the LHC results, we also provide results for the Tevatron.

\begin{table}[]
\begin{tabular}{|c|r|r|r|r|}
\hline
\multirow{2}{*}{}     & \multicolumn{3}{c|}{PMC}                                                                                                    & \multicolumn{1}{c|}{~~~Conv.~~~} \\ \cline{2-5}
                      & \multicolumn{1}{c|}{($q_2, q_3, q_3$)} & \multicolumn{1}{c|}{($q_2, q_3, m_t$)} & \multicolumn{1}{c|}{($q'_2, q'_3, q'_3$)} & \multicolumn{1}{c|}{$m_t$} \\ \hline
$\alpha_s^2$ {[}pb{]} & $405.7$ & $405.7$ & $405.4$ & $406.9$ \\
$\alpha_s^3$ {[}pb{]} & $256.4$ & $256.4$ & $256.7$ & $220.8$ \\
$\alpha_s^4$ {[}pb{]} & $76.4$ & $53.8$ & $76.4$ & $81.5$ \\ \hline
NNLO {[}pb{]} & $738.4$ & $715.9$ & $738.5$ & $709.2^{+28.1}_{-37.2}$ \\ \hline
\end{tabular}
\caption{The $gg$ channel's contribution to the LHC13 cross--section for various PMC scale choices.}
\label{tab:ambiguities_gg}
\end{table}

\begin{table}[]
\begin{tabular}{|c|r|r|r|r|}
\hline
\multirow{2}{*}{}     & \multicolumn{3}{c|}{PMC}                                                                                                    & \multicolumn{1}{c|}{~~~Conv.~~~} \\ \cline{2-5}
                      & \multicolumn{1}{c|}{($q_2, q_3, q_3$)} & \multicolumn{1}{c|}{($q_2, q_3, m_t$)} & \multicolumn{1}{c|}{($q'_2, q'_3, q'_3$)} & \multicolumn{1}{c|}{$m_t$} \\ \hline
$\alpha_s^2$ {[}pb{]} & $62.4$ & $62.4$ & $65.1$ & $68.5$ \\
$\alpha_s^3$ {[}pb{]} & $41.7$ & $41.7$ & $28.4$ & $8.5$ \\
$\alpha_s^4$ {[}pb{]} & $-32.3$ & $-5.2$ & $-14.8$ & $4.7$ \\ \hline
NNLO {[}pb{]} & $71.8$ & $98.9$ & $78.7$ & $81.8^{+1.9}_{-2.2}$ \\ \hline
\end{tabular}
\caption{As in Table~\ref{tab:ambiguities_gg} but for the $q\bar{q}$ channel.}
\label{tab:ambiguities_qqbar}
\end{table}

\begin{table}[]
\centering
\begin{tabular}{|l|l|l|}
\hline
\multicolumn{1}{|c}{} & \multicolumn{1}{|c}{~~~LHC13~~~} & \multicolumn{1}{|c|}{~~~Tevatron~~~} \\
\hline
$\sigma_{\textrm{PMC}}[q_2, q_3, q_3]$ & $813$ & $6.48$ \\
$\sigma_{\textrm{PMC}}[q_2, q_3, m_t]$ & $818$ & $8.30$ \\
$\sigma_{\textrm{PMC}}[q'_2, q'_3, q'_3]$ & $820$ & $6.97$ \\
$\sigma_{\textrm{Conventional}}[m_t]$ & $794^{+28}_{-39}$ & $7.06^{+0.21}_{-0.25}$ \\
\hline $\sigma_{\textrm{Experimental}}$ & $\begin{aligned} 818 &\pm 36 \textsc{ [atlas]} \\ 803 &\pm 32 \textsc{ [cms]}\end{aligned}$ & $7.60 \pm 0.41$ \\
\hline
\end{tabular}
\caption{Total hadronic cross-section (including all partonic channels) through NNLO.}
\label{table:NNLO_full_cross-sections}
\end{table}

The ambiguity over whether to choose the scales $q_n$ or the scales $q'_n$ has effects that vary in size between partonic channels. In the $gg$ channel, where $q'_n\approx q_n$ (c.f. Fig.~\ref{fig:gg_q23}), the scales $q_n'$ produce similar results to the scales $q_n$, as can be seen in Table~\ref{tab:ambiguities_gg}. In the $q\bar{q}$ channel, however, the scales $q_n'$ differ more substantially from $q_n$ (c.f. Fig.~\ref{fig:qqbar_q23}) and the impact on the cross-section is therefore larger, as shown in Table~\ref{tab:ambiguities_qqbar}. The effect of this ambiguity on the overall cross-section, shown in Table~\ref{table:NNLO_full_cross-sections}, is therefore more significant at the Tevatron (where the $q\bar{q}$ channel dominates) than at the LHC (where the $gg$ channel dominates).

The ambiguity over the choice of $q_4$ has a large impact on the value of the cross-section in both of the dominant partonic channels (see Tables~\ref{tab:ambiguities_gg} and~\ref{tab:ambiguities_qqbar}). We note that the numerical impact of the choice of $q_4$ on the BLM/PMC predictions is comparable to that of the choice of $\mu_R$ on the conventional predictions. When the contributions from all partonic channels are combined into a hadron-level cross-section (Table~\ref{table:NNLO_full_cross-sections}), the effect of the $q_4$ ambiguity somewhat cancels between channels in the LHC cross-section, but is significantly larger in the Tevatron cross-section. In principle, the BLM/PMC method does prescribe a value for $q_4$, but it requires information from the currently unknown N\textsuperscript{3}LO cross-section. Note, however, that a new arbitrary scale, $q_5$, would appear at N\textsuperscript{3}LO --- any calculation using the BLM/PMC method will always involve one arbitrary scale.

\subsection{Comparison of strategies to handle the $q_4$ ambiguity}\label{subsec:NLO_analysis}

It was found in the previous section that the ambiguity over the highest-order scale, $q_4$, has a significant impact on the prediction for the cross-section. In the literature describing the BLM/PMC method, it is suggested \cite{Brodsky:2011ta, Brodsky:2013vpa} that the ambiguity could have been resolved if we had information from the next perturbative order in $\as$. In this section, we will explore 4 ways of handling the ambiguity over the highest-order scale, including the suggested approach of ``peeking'' at the next perturbative order. We choose to work with the NLO cross-section, allowing us the possibility to ``peek'' at the NNLO cross-section when setting the PMC scales.

Only 2 scales appear in the NLO cross-section: $q_2$ and $q_3$. At this order in perturbation theory, $q_3$ is arbitrary since it relies on information appearing in the NNLO cross-section (c.f. Eqs.~\ref{eq:logQ3} and~\ref{eq:logQ3prime}). We will calculate the NLO cross-section while exploring the following possible choices for the PMC scales ($q_2$, $q_3$):
\begin{enumerate}
\item ($q'_2$, $q'_2$) Of the PMC scales defined in Section~\ref{sec:BLM/PMC}, the only one that does not require information from the NNLO calculation is the scale $q'_2$ (c.f. Eq.~\ref{eq:logQ2prime}). (In fact, $q'_2$ was the scale prescribed in the original BLM paper \cite{Brodsky:1982gc}.) Hence, one option is to set both PMC scales to be $q'_2$.
\item ($q'_2$, $m_t$) To explore the impact of the arbitrary scale $q_3$ without relying on any NNLO information, we can set $q_3 = m_t$ and compare against the results of the previous scale choice.
\item ($q'_2$, $q'_3$) If we allow ourselves to peek at the NNLO cross-section, we can use the full NNLO PMC scales $q'_2$ and $q'_3$ defined in Eqs.~\ref{eq:logQ2prime} and~\ref{eq:logQ3prime}.
\item ($q_2$, $q_3$) Alternatively, again peeking at the NNLO cross-section, we could choose to use the scales $q_n$ (defined in Eqs.~\ref{eq:logQ2} and~\ref{eq:logQ3}) rather than $q'_n$.
\end{enumerate}

The resulting contributions of the $gg$ and $q\bar{q}$ channels to the LHC cross-section are shown in Tables~\ref{tab:NLO_gg} and~\ref{tab:NLO_qqbar} respectively. The total NLO cross-section, incorporating the contributions from all partonic channels, is shown in Table~\ref{table:NLO_full_cross-sections}, where alongside the LHC results, we also provide results for the Tevatron.

\begin{table}[]
\begin{tabular}{|c|r|r|r|r|r|}
\hline
\multirow{2}{*}{} & \multicolumn{4}{c|}{PMC} & \multicolumn{1}{c|}{Conv.} \\ \cline{2-6}
 & \multicolumn{1}{c|}{($q'_2, q'_2$)} & \multicolumn{1}{c|}{($q'_2, m_t$)} & \multicolumn{1}{c|}{($q'_2, q'_3$)} & \multicolumn{1}{c|}{($q_2, q_3$)} & \multicolumn{1}{c|}{$m_t$} \\ \hline
$\alpha_s^2$ {[}pb{]} & $405.4$ & $405.4$ & $405.4$ & $405.7$ & $406.9$ \\
$\alpha_s^3$ {[}pb{]} & $221.3$ & $222.3$ & $256.7$ & $256.4$ & $220.8$ \\ \hline
NLO {[}pb{]} & $626.7$ & $627.7$ & $662.1$ & $662.1$ & $627.7^{+67.6}_{-63.6}$ \\ \hline
\end{tabular}
\caption{The $gg$ channel's contribution to the LHC13 cross--section at NLO using various scale choices}
\label{tab:NLO_gg}
\end{table}

\begin{table}[]
\begin{tabular}{|c|r|r|r|r|r|}
\hline
\multirow{2}{*}{} & \multicolumn{4}{c|}{PMC} & \multicolumn{1}{c|}{Conv.} \\ \cline{2-6}
 & \multicolumn{1}{c|}{($q'_2, q'_2$)} & \multicolumn{1}{c|}{($q'_2, m_t$)} & \multicolumn{1}{c|}{($q'_2, q'_3$)} & \multicolumn{1}{c|}{($q_2, q_3$)} & \multicolumn{1}{c|}{$m_t$} \\ \hline
$\alpha_s^2$ {[}pb{]} & $65.1$ & $65.1$ & $65.1$ & $62.4$ & $68.5$ \\
$\alpha_s^3$ {[}pb{]} & $11.2$ & $12.2$ & $28.4$ & $41.7$ & $8.5$ \\ \hline
NLO {[}pb{]} & $76.3$ & $77.2$ & $93.5$ & $104.1$ & $77.0^{+1.3}_{-3.9}$ \\ \hline
\end{tabular}
\caption{As in Table~\ref{tab:NLO_gg} but for the $q\bar{q}$ channel.}
\label{tab:NLO_qqbar}
\end{table}

Comparing the choices $(q'_2, q'_2)$, $(q'_2, m_t)$, and $(q'_2, q'_3)$, one sees that the numerical impact of the choice of $q_3$ on the PMC prediction can be similar to the impact of the choice of $\mu_R$ on the conventional prediction. This is analogous to the findings of the previous section in relation to the $q_4$ ambiguity at NNLO.

We note that the scale choices $(q'_2, q'_2)$ and $(q'_2, m_t)$ --- obtained using only information available at NLO --- lead to very different cross-sections compared to the scale choices $(q'_2, q'_3)$ and $(q_2, q_3)$, which were obtained by peeking at the next perturbative order. In other words, when handling the ambiguous highest-order PMC scale, the two approaches appearing in the literature (to either use an existing PMC scale or instead peek at the next perturbative order) yield very different results to one another, as well as to other plausible choices for this scale. The arbitrary choice of a value for the highest-order PMC scale thus remains an open problem.

\begin{table}[]
\centering
\begin{tabular}{|l|l|l|}
\hline
\multicolumn{1}{|c}{} & \multicolumn{1}{|c}{~~~LHC13~~~} & \multicolumn{1}{|c|}{~~~Tevatron~~~} \\
\hline
$\sigma_{\textrm{PMC}}[q'_2, q'_2]$ & $709$ & $6.52$ \\
$\sigma_{\textrm{PMC}}[q'_2, m_t]$ & $711$ & $6.51$ \\
$\sigma_{\textrm{PMC}}[q'_2, q'_3]$ & $762$ & $7.86$ \\
$\sigma_{\textrm{PMC}}[q_2, q_3]$ & $773$ & $8.59$ \\
$\sigma_{\textrm{Conventional}}[m_t]$ & $711^{+71}_{-69}$ & $6.51^{+0.30}_{-0.44}$ \\
\hline $\sigma_{\textrm{Experimental}}$ & $\begin{aligned} 818 &\pm 36 \textsc{ [atlas]} \\ 803 &\pm 32 \textsc{ [cms]}\end{aligned}$ & $7.60 \pm 0.41$ \\
\hline
\end{tabular}
\caption{Total hadronic cross-section (including all partonic channels) through NLO. For NNLO, see Table~\ref{table:NNLO_full_cross-sections}.}
\label{table:NLO_full_cross-sections}
\end{table}

\section{Conclusion}

The BLM/PMC procedure is a proposed method for eliminating the renormalisation scale ambiguity in perturbative QCD. In this work, we have presented three ambiguities in the BLM/PMC procedure itself. We have studied these ambiguities using the example of $t\bar{t}$ production at NNLO in QCD and have found two of the ambiguities to have a significant numerical impact on the computed cross-sections.

One of these ambiguities lies in the definition of the PMC scales $q_n$: we give an example of an alternative set of scales, $q'_n$, which satisfy the PMC requirement that terms proportional to the QCD $\beta$-function coefficients are to be absorbed into the running coupling. The other ambiguity arises because in any calculation employing the BLM/PMC scale-setting procedure, the highest-order scale (in this case, $q_4$) remains arbitrary. We find the numerical impact of each of these ambiguities to be comparable to the impact of the choice of $\mu_R$ in the conventional scale-setting approach.

In the existing literature on the BLM/PMC method, it is asserted that the $q_4$ ambiguity could in principle be resolved using information from even higher perturbative orders, and that it should otherwise by handled by using an existing PMC scale. We find that the cross-sections arising within these two approaches can differ markedly from one another, as well as from the cross-sections arising from other plausible choices for this scale.

In summary, while the BLM/PMC procedure is well-motivated, it contains important ambiguities with significant numerical impact on the predicted values for physical observables. We also emphasise that even an unambiguous scale-setting prescription would not remove the theoretical uncertainties in physical predictions, since these uncertainties ultimately arise from missing higher orders in $\as$. We hope our work will lead to an improved understanding of the problem of scale settings which, in turn, should result in improved theoretical predictions for hadron collider processes.

\begin{acknowledgments}
A.M. thanks the Department of Physics at Princeton University for hospitality during the completion of this work. This project has received funding from the European Research Council (ERC) under the European Union's Horizon 2020 research and innovation programme (grant agreement No 683211). This work is also supported by the UK STFC grants ST/L002760/1 and ST/K004883/1. 
\end{acknowledgments}


\begin{thebibliography}{99}

\bibitem{Nason:1989zy} 
  P.~Nason, S.~Dawson and R.~K.~Ellis,
  Nucl.\ Phys.\ B {\bf 327}, 49 (1989)
  Erratum: [Nucl.\ Phys.\ B {\bf 335}, 260 (1990)].

\bibitem{Beenakker:1990maa} 
  W.~Beenakker, W.~L.~van Neerven, R.~Meng, G.~A.~Schuler and J.~Smith,
  Nucl.\ Phys.\ B {\bf 351}, 507 (1991).

\bibitem{Mangano:1991jk} 
  M.~L.~Mangano, P.~Nason and G.~Ridolfi,
  Nucl.\ Phys.\ B {\bf 373}, 295 (1992).
  
\bibitem{Catani:2001cc} 
  S.~Catani, F.~Krauss, R.~Kuhn and B.~R.~Webber,
  JHEP {\bf 0111}, 063 (2001)
  [hep-ph/0109231].

\bibitem{Maltoni:2003pn} 
  F.~Maltoni, Z.~Sullivan and S.~Willenbrock,
  Phys.\ Rev.\ D {\bf 67}, 093005 (2003)
  [hep-ph/0301033].

\bibitem{Boos:2003yi} 
  E.~Boos and T.~Plehn,
  Phys.\ Rev.\ D {\bf 69}, 094005 (2004)
  [hep-ph/0304034].

\bibitem{Alwall:2007fs} 
  J.~Alwall {\it et al.},
  Eur.\ Phys.\ J.\ C {\bf 53}, 473 (2008)
  [arXiv:0706.2569 [hep-ph]].

\bibitem{Bauer:2009km} 
  C.~W.~Bauer and B.~O.~Lange,
  arXiv:0905.4739 [hep-ph].

\bibitem{Melnikov:2009wh} 
  K.~Melnikov and G.~Zanderighi,
  Phys.\ Rev.\ D {\bf 81}, 074025 (2010)
  [arXiv:0910.3671 [hep-ph]].

\bibitem{Ahrens:2010zv} 
  V.~Ahrens, A.~Ferroglia, M.~Neubert, B.~D.~Pecjak and L.~L.~Yang,
  JHEP {\bf 1009}, 097 (2010)
  [arXiv:1003.5827 [hep-ph]].
  
\bibitem{Berger:2010zx} 
  C.~F.~Berger {\it et al.},
  Phys.\ Rev.\ Lett.\  {\bf 106}, 092001 (2011)
  [arXiv:1009.2338 [hep-ph]].
    
\bibitem{Denner:2012yc} 
  A.~Denner, S.~Dittmaier, S.~Kallweit and S.~Pozzorini,
  JHEP {\bf 1210}, 110 (2012)
  [arXiv:1207.5018 [hep-ph]].

\bibitem{Anastasiou:2016cez} 
  C.~Anastasiou, C.~Duhr, F.~Dulat, E.~Furlan, T.~Gehrmann, F.~Herzog, A.~Lazopoulos and B.~Mistlberger,
  JHEP {\bf 1605}, 058 (2016)
  [arXiv:1602.00695 [hep-ph]].
  
\bibitem{Boughezal:2016yfp} 
  R.~Boughezal, X.~Liu and F.~Petriello,
  Phys.\ Lett.\ B {\bf 760}, 6 (2016)
  [arXiv:1602.05612 [hep-ph]].

\bibitem{Czakon:2016dgf} 
  M.~Czakon, D.~Heymes and A.~Mitov,
  JHEP {\bf 1704}, 071 (2017)
  [arXiv:1606.03350 [hep-ph]].

\bibitem{Currie:2017eqf} 
  J.~Currie, A.~Gehrmann-De Ridder, T.~Gehrmann, E.~W.~N.~Glover, A.~Huss and J.~Pires,
  Phys.\ Rev.\ Lett.\  {\bf 119}, no. 15, 152001 (2017)
  [arXiv:1705.10271 [hep-ph]].

\bibitem{Czakon:2018nun} 
  M.~Czakon, A.~Ferroglia, D.~Heymes, A.~Mitov, B.~D.~Pecjak, D.~J.~Scott, X.~Wang and L.~L.~Yang,
  JHEP {\bf 1805}, 149 (2018)
  [arXiv:1803.07623 [hep-ph]].

\bibitem{Brodsky:1982gc} 
  S.~J.~Brodsky, G.~P.~Lepage and P.~B.~Mackenzie,
  Phys.\ Rev.\ D {\bf 28}, 228 (1983).

\bibitem{Brodsky:2011ig} 
  S.~J.~Brodsky and L.~Di Giustino,
  Phys.\ Rev.\ D {\bf 86}, 085026 (2012)
  [arXiv:1107.0338 [hep-ph]].

\bibitem{Brodsky:2011ta}
  S.~J.~Brodsky and X.~G.~Wu,
  Phys.\ Rev.\ D {\bf 85} (2012) 034038
   Erratum: [Phys.\ Rev.\ D {\bf 86} (2012) 079903]
  [arXiv:1111.6175 [hep-ph]].

\bibitem{Spira:1995rr} 
  M.~Spira, A.~Djouadi, D.~Graudenz and P.~M.~Zerwas,
  Nucl.\ Phys.\ B {\bf 453}, 17 (1995)
  [hep-ph/9504378].

\bibitem{Belitsky:2001nq} 
  A.~V.~Belitsky and D.~Mueller,
  Phys.\ Lett.\ B {\bf 513}, 349 (2001)
  [hep-ph/0105046].

\bibitem{Belitsky:2005qn} 
  A.~V.~Belitsky and A.~V.~Radyushkin,
  Phys.\ Rept.\  {\bf 418}, 1 (2005)
  [hep-ph/0504030].

\bibitem{Ivanov:2004zv} 
  D.~Y.~Ivanov, L.~Szymanowski and G.~Krasnikov,
  JETP Lett.\  {\bf 80}, 226 (2004)
  [Pisma Zh.\ Eksp.\ Teor.\ Fiz.\  {\bf 80}, 255 (2004)]
  Erratum: [JETP Lett.\  {\bf 101}, no. 12, 844 (2015)]
  [hep-ph/0407207].

\bibitem{Anikin:2004jb} 
  I.~V.~Anikin, B.~Pire, L.~Szymanowski, O.~V.~Teryaev and S.~Wallon,
  Eur.\ Phys.\ J.\ C {\bf 42}, 163 (2005)
  [hep-ph/0411408].

\bibitem{Brodsky:1997dh} 
  S.~J.~Brodsky, C.~R.~Ji, A.~Pang and D.~G.~Robertson,
  Phys.\ Rev.\ D {\bf 57}, 245 (1998)
  [hep-ph/9705221].

\bibitem{Bakulev:2000uh} 
  A.~P.~Bakulev, A.~V.~Radyushkin and N.~G.~Stefanis,
  Phys.\ Rev.\ D {\bf 62}, 113001 (2000)
  [hep-ph/0005085].

\bibitem{Voloshin:1995sf} 
  M.~B.~Voloshin,
  Int.\ J.\ Mod.\ Phys.\ A {\bf 10}, 2865 (1995)
  [hep-ph/9502224].

\bibitem{Bigi:1997fj} 
  I.~I.~Y.~Bigi, M.~A.~Shifman and N.~Uraltsev,
  Ann.\ Rev.\ Nucl.\ Part.\ Sci.\  {\bf 47}, 591 (1997)
  [hep-ph/9703290].

\bibitem{Brodsky:2012rj} 
  S.~J.~Brodsky and X.~G.~Wu,
  Phys.\ Rev.\ Lett.\  {\bf 109}, 042002 (2012)
  [arXiv:1203.5312 [hep-ph]].

\bibitem{Brodsky:2012sz} 
  S.~J.~Brodsky and X.~G.~Wu,
  Phys.\ Rev.\ D {\bf 86}, 014021 (2012)
  Erratum: [Phys.\ Rev.\ D {\bf 87}, no. 9, 099902 (2013)]
  [arXiv:1204.1405 [hep-ph]].

\bibitem{Brodsky:2012ik} 
  S.~J.~Brodsky and X.~G.~Wu,
  Phys.\ Rev.\ D {\bf 85}, 114040 (2012)
  [arXiv:1205.1232 [hep-ph]].

\bibitem{Wang:2014sua} 
  S.~Q.~Wang, X.~G.~Wu, Z.~G.~Si and S.~J.~Brodsky,
  Phys.\ Rev.\ D {\bf 90}, no. 11, 114034 (2014)
  [arXiv:1410.1607 [hep-ph]].

\bibitem{Wang:2015lna} 
  S.~Q.~Wang, X.~G.~Wu, Z.~G.~Si and S.~J.~Brodsky,
  Phys.\ Rev.\ D {\bf 93}, no. 1, 014004 (2016)
  [arXiv:1508.03739 [hep-ph]].

\bibitem{Wang:2017kyd} 
  S.~Q.~Wang, X.~G.~Wu, Z.~G.~Si and S.~J.~Brodsky,
  Eur.\ Phys.\ J.\ C {\bf 78}, no. 3, 237 (2018)
  [arXiv:1703.03583 [hep-ph]].

\bibitem{Kataev:2014jba}
  A.~L.~Kataev and S.~V.~Mikhailov,
  Phys.\ Rev.\ D {\bf 91} (2015) no.1,  014007
  [arXiv:1408.0122 [hep-ph]].

\bibitem{Kataev:2014zwa}
  A.~L.~Kataev,
  J.\ Phys.\ Conf.\ Ser.\  {\bf 608} (2015) no.1,  012078
  [arXiv:1411.2257 [hep-ph]].

\bibitem{Kataev:2016aib}
  A.~L.~Kataev and S.~V.~Mikhailov,
  JHEP {\bf 1611} (2016) 079
  [arXiv:1607.08698 [hep-th]].

\bibitem{Aleshin:2019yqj}
  S.~S.~Aleshin, A.~L.~Kataev and K.~V.~Stepanyantz,
  JHEP {\bf 1903} (2019) 196
  [arXiv:1902.08602 [hep-th]].

\bibitem{Garkusha:2018mua}
  A.~V.~Garkusha, A.~L.~Kataev and V.~S.~Molokoedov,
  JHEP {\bf 1802} (2018) 161
  [arXiv:1801.06231 [hep-ph]].

\bibitem{Baernreuther:2012ws} 
  P.~Baernreuther, M.~Czakon and A.~Mitov,
  Phys.\ Rev.\ Lett.\  {\bf 109}, 132001 (2012)
  [arXiv:1204.5201 [hep-ph]].
  
\bibitem{Czakon:2012zr} 
  M.~Czakon and A.~Mitov,
  JHEP {\bf 1212}, 054 (2012)
  [arXiv:1207.0236 [hep-ph]].
  
\bibitem{Czakon:2012pz} 
  M.~Czakon and A.~Mitov,
  JHEP {\bf 1301}, 080 (2013)
  [arXiv:1210.6832 [hep-ph]].

\bibitem{Czakon:2013goa} 
  M.~Czakon, P.~Fiedler and A.~Mitov,
  Phys.\ Rev.\ Lett.\  {\bf 110}, 252004 (2013)
  [arXiv:1303.6254 [hep-ph]].

\bibitem{Czakon:2014xsa} 
  M.~Czakon, P.~Fiedler and A.~Mitov,
  Phys.\ Rev.\ Lett.\  {\bf 115}, no. 5, 052001 (2015)
  [arXiv:1411.3007 [hep-ph]].

\bibitem{Brodsky:2013vpa} 
  S.~J.~Brodsky, M.~Mojaza and X.~G.~Wu,
  Phys.\ Rev.\ D {\bf 89}, 014027 (2014)
  [arXiv:1304.4631 [hep-ph]].
  
\bibitem{Beneke:2009ye} 
  M.~Beneke, M.~Czakon, P.~Falgari, A.~Mitov and C.~Schwinn,
  Phys.\ Lett.\ B {\bf 690}, 483 (2010)
  [arXiv:0911.5166 [hep-ph]].

\bibitem{Beneke:2016jpx} 
  M.~Beneke and P.~Ruiz-Femenia,
  JHEP {\bf 1608}, 145 (2016)
  [arXiv:1606.02434 [hep-ph]].
  
\bibitem{Ball:2017nwa} 
  R.~D.~Ball {\it et al.} [NNPDF Collaboration],
  Eur.\ Phys.\ J.\ C {\bf 77}, no. 10, 663 (2017)
  [arXiv:1706.00428 [hep-ph]].

\bibitem{Dulat:2015mca} 
  S.~Dulat {\it et al.},
  Phys.\ Rev.\ D {\bf 93}, no. 3, 033006 (2016)
  [arXiv:1506.07443 [hep-ph]].

\bibitem{Czakon:2011xx} 
  M.~Czakon and A.~Mitov,
  Comput.\ Phys.\ Commun.\  {\bf 185}, 2930 (2014)
  [arXiv:1112.5675 [hep-ph]].
  
\bibitem{Aaboud:2016pbd}
  M.~Aaboud {\it et al.} [ATLAS Collaboration],
  Phys.\ Lett.\ B {\bf 761} (2016) 136
   Erratum: [Phys.\ Lett.\ B {\bf 772} (2017) 879]
  [arXiv:1606.02699 [hep-ex]].

\bibitem{Sirunyan:2018goh} 
  A.~M.~Sirunyan {\it et al.} [CMS Collaboration],
  Eur.\ Phys.\ J.\ C {\bf 79}, no. 5, 368 (2019)
  [arXiv:1812.10505 [hep-ex]].

\bibitem{Aaltonen:2013wca} 
  T.~A.~Aaltonen {\it et al.} [CDF and D0 Collaborations],
  Phys.\ Rev.\ D {\bf 89}, no. 7, 072001 (2014)
  [arXiv:1309.7570 [hep-ex]].

\bibitem{Grunberg:1991ac}
  G.~Grunberg and A.~L.~Kataev,
  Phys.\ Lett.\ B {\bf 279} (1992) 352.

\bibitem{Cvetic:2016rot}
  G.~Cvetic and A.~L.~Kataev,
  Phys.\ Rev.\ D {\bf 94} (2016) no.1,  014006
  [arXiv:1604.00509 [hep-ph]].

\end{thebibliography}
\end{document}